\documentclass{aa}
\usepackage{epsfig}
\usepackage{astron}
\newcommand*{\figun}{
\begin{figure}[ht!]
\epsfig{file=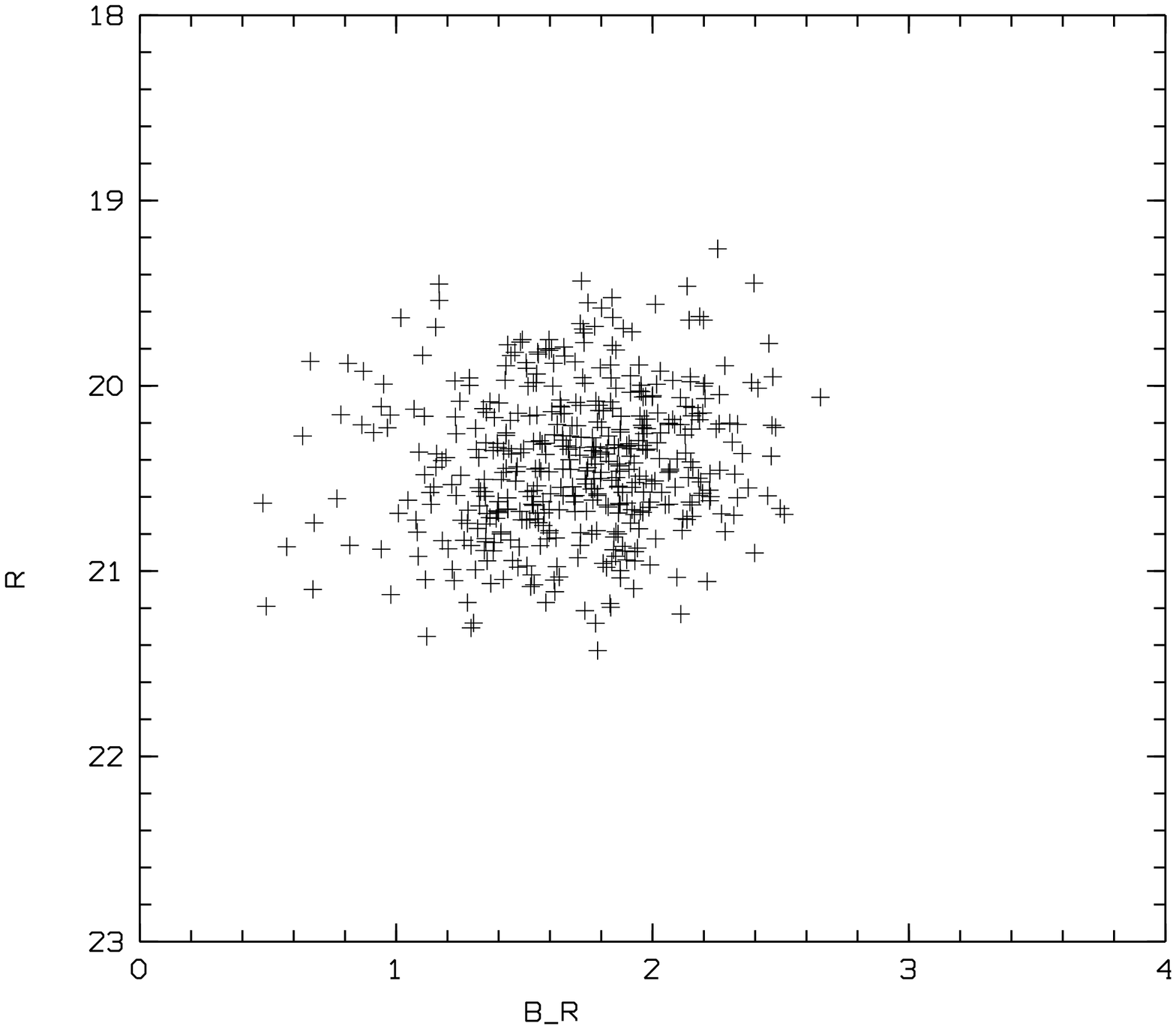,width=.42\textwidth,clip=,angle=-90}
\caption[]{Colour Magnitude diagram at observed maximum brightness 
for the 459 variable stars which have one B measurement. 
\label{figun}}
\end{figure}}

\newcommand{\beq}{\begin{equation}}
\newcommand{\beqa}{\begin{eqnarray}}
\newcommand{\eeq}{\end{equation}}
\newcommand{\eeqa}{\end{eqnarray}}

\begin{document}
\bibliographystyle{astron}


\title{Variable stars towards the bulge of M31: the AGAPE catalogue
\thanks{Complete Table 3 is available in electronic form at the CDS
  via anonymous ftp to cdsarc.u-strasbg.fr (130.79.128.5) or via
  http://cdsweb.u-strasbg.fr/cgi-bin/qcat?J/A+A/}
\thanks{Based on data collected with the 2\,m Bernard Lyot Telescope (TBL)
operated by INSU-CNRS and Pic-du-Midi Observatory (USR 5026).
\protect\newline The experiment was funded by IN2P3 and INSU of CNRS}}

\author{ R. Ansari \inst{1} \and M. Auri{\`e}re \inst{2} \and P. Baillon \inst{3
}
\and A. Bouquet \inst{4} \and G. Coupinot \inst{2} \and Ch. Coutures \inst{5}
 \and C. Ghesqui{\`e}re \inst{4} \and Y.~Giraud-H{\'e}raud \inst{4} \and
D. Gillieron \inst{2} \and P. Gondolo \inst{6}  \and J. Hecquet \inst{7} \thanks{deceased} 
\and J. Kaplan \inst{4} \and
A. Kim \inst{4,8} \and Y. Le Du \inst{4} \and A.L. Melchior \inst{9}
\and M. Moniez \inst{1} \and J.P. Picat \inst{7} \and G. Soucail 
\inst{7}}

\offprints{M. Auri{\`e}re, michel.auriere@obs-mip.fr}

\institute{Laboratoire de l'Acc{\'e}l{\'e}rateur Lin{\'e}aire, Universit{\'e} Paris-Sud,
91405 Orsay, France 
\and Observatoire Midi-Pyr{\'e}n{\'e}es, unit{\'e} associ{\'e}e au CNRS (UMR 5572),
57 Avenue d'Azereix, BP 826, 65008 Tarbes Cedex, France 
\and CERN, 1211 Gen{\`e}ve 23, Switzerland
\and Physique Corpusculaire et Cosmologie, Coll{\`e}ge de France, laboratoire
associ{\'e} au  CNRS-IN2P3 (UMR 7550), 11 place Marcelin Berthelot, 75231
Paris Cedex 05, France 
\and DAPNIA/SPP, CEN Saclay, 91191 Gif-sur-Yvette, France
\and Department of Physics, University of Utah, 115 S 1400 E Rm 201, Salt Lake City, UT 84112-0830, USA
\and Observatoire Midi-Pyr{\'e}n{\'e}es, unit{\'e} associ{\'e}e au CNRS (UMR 5572), 14 avenue Belin, 31400 Toulouse, France
\and Lawrence Berkeley National Laboratory, Berkeley, CA 94720,USA
\and LERMA, Observatoire de Paris, 61 avenue de l'Observatoire, 75014
Paris, France}

\date{submitted}

\abstract{
We present the AGAPE  astrometric and photometric
catalogue of 1579 variable stars in a $14' \times 10'$ field centred on
M31. This work  is the first survey devoted to variable stars in the
bulge of M31. 
The $R$ magnitudes
of the objects and the $B-R$ colours  suggest that our sample is
dominated by red long-period variable stars (LPV), with a possible overlap with
Cepheid-like type II stars. Fits of the light curves with sinusoids
suggest that a large fraction of the stars correspond to periodic or
semi-periodic objects with periods longer than 100 days. Twelve nova candidates are identified. Correlations with other catalogues suggest that 2 novae could be recurrent novae and provide possible optical counterparts for 2 supersoft X-ray sources candidates observed with Chandra. 
}

\maketitle

\section{Introduction} 

AGAPE \cite{AGAPE1} made a survey
of several fields in the central part of M31 to look for
gravitational microlensing events. As a major by-product, this
survey discovered 1579 variable objects in a $14' \times 10'$
field centred on M31. The aim of this paper is to catalogue these
objects.
Due to the high star
background, pixel lensing techniques have to be used to detect flux
variations of individual 
stars. Since we are observing brightness fluctuations of unresolved stars, variable
stars, particularly those that are intrinsically faint, are a
significant background in the detection of real microlensing events.
On the other hand, variable stars are also of high interest in their
own right. Systematic surveys looking for variable stars in M31 have been
first done by Hubble \cite*{hubble29}and Baade and Swope  \cite*{baade63,baade65}.
  Recent  surveys on  large fields have
been used to search for Cepheids \cite{magnier97} and
cosmological calibrators (DIRECT project \cite{kaluzny98}).
However, except for novae and X-ray sources Kaaret \cite*{kaaret02}, Kong et al. \cite*{kong02},  published work avoid the bulge region.  
Our study is then the first one done  on this central zone and it leads to the
identification of some  variable stars poorly known in M31: Miras and
long period Cepheid-like objects.
 Our catalogue
is drawn from 3-seasons of measurements and our field has significant intersection with those of other studies (\cite{crotts96}, DIRECT field E, \cite{riffeser01}) and the Point AGAPE collaboration using the INT \cite{kerins01}.

Sect. 2 presents AGAPE observations and variable selection. Sect. 3 presents the astrometry 
as well as our photometry and period search (Time Series Analysis).  Sect. 4 discusses
our type-identified variable stars and sect. 5 presents our complete 
catalogue as well as some conclusions.

\section{AGAPE observations and variable selection}

\subsection{AGAPE observations and pixel method photometry}

The AGAPE observations were made at the 2\,m Bernard Lyot telescope (TBL) of
the Pic du Midi Observatory with the f/8 spectro-reducer ISARD. A
thin Tektronik $1024\times1024$ CCD was used with a useful field of $4' \times 4.5'$
with $0.3''$ pixels.  Since the field of ISARD is small, 6 fields
(called A, B, C, D, E, F) were  necessary to cover one $14' \times 10'$ field
centred on M31 and oriented along its main axis. The exposure
times were generally  30(20)\,min in the $B(R)$ passband. An additional
field, Z, centred on the nucleus of M31 and with large overlap with A and B fields, was taken at the beginning of
each night, as a reference to help in the pointing of the
telescope. The exposures were as short as 1\,min in both the $B$ and $R$
passbands for the Z central field. Since the short exposures prevent
detection of faint or small amplitude variable objects, only novae
are presented for the Z field data in this paper. The observing campaign
ran from 1994 to 1996. It turned out that it was impossible to monitor
all the fields in both colours each night. We put a
priority on the first four fields, with an emphasis on red
exposures.  We obtained 69
observing nights in $R$: 19 in 1994, 43 in 1995 and 7 in
1996. The number of images taken in each field during the whole survey
is given in table \ref{timages}.

\begin{table}

\begin{center}
\begin{tabular}{ccccc}
\hline
\hline
year & 1994 & 1995 & 1996 & total \\

filter & $R/B$ & $R/B$ & $R/B$ & $R/B$ \\
\hline
A & 25/10 & 51/21 & 8/2 & 84/33 \\

B & 24/10 & 41/21 & 7/2 & 73/33 \\

C & 20/8 & 40/16 & 7/0 & 67/24 \\

D & 18/5 & 34/14 & 6/0 & 59/19 \\

E & 6/0 & 34/10 & 6/0 & 46/10 \\

F & 5/0 & 27/8 & 6/0 & 38/8 \\

Z & 32/12 & 55/20 & 13/1 & 100/33 \\
\hline
\end{tabular}
\end{center} 
\caption[]{Number of images taken during the AGAPE observations}
\label{timages}
\end{table}

The AGAPE detection procedure is described in Ansari et
al. \cite*{AGAPE1}. It is based on super-pixel (7x7 pixels) photometry 
with sides roughly twice the standard seeing. 
These super-pixels are photometrically normalised to a
reference frame and corrected for seeing variations. Then we
transform the super-pixel ADU photometry to Johnson-Cousins $B$ and $R$
magnitudes as described in sect. 3.2.

\subsection{Variable stars selection}

During the selection procedure, we excluded an area of 20 pixels
around well-resolved stars. Automatic procedures are efficient to detect stars brighter than $R=19$. \cite{ledu00} studies the relation of this influence zone with seeing. These resolved stars should contain
no or very few M31 variable stars but stars from our Galaxy and M31
globular clusters.  The light curves for each uncontaminated  super-pixel are analysed. Baseline for background
flux is taken as the minimum of a sliding average calculated on 4 to 8 
consecutive
points depending of the time sampling of the field. 
One bump is defined as beginning when 3 consecutive points lie
$3 \sigma$ above background and ending when 2 consecutive points fall
below $3 \sigma$. For each bump, we compute its likelihood function :

$$ L = -\ln \left( \prod_{i \in {\mathrm bump}} P(\Phi>\Phi_i \mid <\Phi>_i,\sigma_i)\right)$$

One superpixel is selected as including a variable object when it
contains one bump with its likelihood function greater than 100. We choose a high value for $L$ in order to eliminate false variable stars. These
criteria lead to the selection of  1579 variable stars. 

\section {Astrometry, photometry and period search}

\subsection{Astrometry}

Astrometry is based upon the catalogue of Magnier et
al. \cite*{magnier92}, the most populous one available towards the bulge of M31
with the most homogeneous coverage of our field.
This catalogue
is connected to the Space Telescope Guide Star Catalog with an accuracy
better than $1''$ \cite{magnier93,haiman94}. For each of
AGAPE fields  B to F  we have from 6 to 20 stars in common with
this catalogue. The final standard deviation on the
positions is in the range $0.10''-0.14''$. However, for the  near-central
fields Z and A, less than 3 stars are identified as common and
we were obliged to use another reference catalogue. We thus used the one
obtained by Auri{\`e}re et al. \cite*{auriere92} for their globular
clusters study, based on the combination of \cite{battistini87} and
\cite{wirth85}. For these two fields the standard deviation on
the distances between our positions and those on the catalogue for 11
and 10 common stars in fields A and Z respectively is
$0.45''$. Finally  we conclude that we have an astrometric 
accuracy in the range  $1''-1.5''$ for our variable star positions, combining our
uncertainties with those of Magnier and the other catalogues.

\figun

\subsection{Photometry}

The super-pixel photometry gives the integrated flux in squares of 7x7
pixels corresponding to 2.1'' x 2.1'' on the sky. We first have to remove the
stellar background which is bright and chaotic in the M31 bulge. Almost all
 variable stars are not resolved nor significantly measured at
 minimum. In this case we can consider the background as being the average
 minimum flux, 
which is taken as 
the minimum of the sliding average on five points along the light curve. 
However, the star always contributes at
some level even at minimum brightness and we
also compute the average flux (with $3 \sigma$  rejection) in a zone
around the object. The background is taken as the lesser of these
two values and is subtracted from flux values. To obtain $B$ and
$R$ Johnson-Cousins magnitudes, we use the colour equations as given in
Ansari et al. \cite*{AGAPE1} transforming between super-pixel
and PSF photometry.
 Since our field contains several ``bright unvariable'' stars measured by Magnier et al. \cite*{magnier92},
we use 36 common stars in $R$ and 19 common stars in $B$ (excluding
central fields A and B) to derive magnitude scale constants for our
data. The standard deviations corresponding to the computation of
these constants are 0.15 and 0.24 for $R$ and $B$ respectively. They are dominated by random scatter and include
our errors and Magnier et al. \cite*{magnier92} errors.
This accuracy is sufficient to characterise the $R$
magnitude at maximum observed brightness of the variable stars and the
colour range for our star sample, but is too poor for using individual
$B-R$ indices for identification purposes. The $B$ magnitude
measurements will be thus used for making a Colour Magnitude Diagram
but will not be given in the final catalogue. 
 To compute the $R$ magnitude
 at maximum we select the 5 brightest $R$ measurements and eliminate the 2 brightest ones to avoid possible residual cosmic rays. $R_{max}$ is the average
of the 3 remaining measurements. To enable a reliable $B$ magnitude measurement and thus a $B-R$ colour, the stars have to emerge significantly from the background. Since there are fewer $B$ measurements, the magnitude at maximum is the average of the 2 brightest measurements. $B-R$ is the difference between the peak $B$ and $R$ in each passband.
Figure 1 gives our Colour Magnitude Diagram at observed maximum
  brightness for the 459 variable stars which have been measured in
  the $B$ passband, in fields A, B, C, D. This diagram is interpreted
in sect. 4.1.

 \begin{table}[tbh]
\begin{center}
\begin{tabular}{cccccc}
\hline
\hline
Ident & $R_{max}$ & \begin{tabular}{c}Period \\(day)\end{tabular} &
Ident & $R_{max}$ & \begin{tabular}{c}Period \\(day)\end{tabular}
\\
\hline
B030&           20.5&   060 &  C032    &       20.6&   187     \\

B098&           21.1&   103 &B005    &       20.3&   188\\

A266&           21.1&   111 &A320    &       20.2&   189\\

B054&           19.9&   116 &A077    &       19.7&   190\\

A058&           19.6&   123&B043    &       20.1&   193\\

C071&           20.9&   125&A189    &       20.4&   194\\

C188&           21.2&   127&A185    &       20.6&   196\\

A014&           20.1&   131&A044    &       20.3&   197\\

C055&           20.5&   132&A162    &       20.0&   197\\

C012&           20.7&   133&A241    &       20.8&   197\\

C081&           20.9&   134&A343&           20.0&   200\\

C013&           20.3&   140&A190&           20.3&   202\\

A059&           20.4&   142&A002&           19.6&   210\\

A018&           20.1&   148&B006&           20.3&   212\\

A046&           20.2&   151&C016&           20.5&   212\\

B014&           20.4&   164&C169&           21.0&   216\\

A286&           20.6&   155&A019&           20.2&   220\\

B058&           20.6&   156&A109&           20.3&   223\\

C017&           20.5&   162&B140&           20.0&   244\\

B165&           20.2&   164&C006&           20.1&   244\\

A079&           19.9&   165&B007&           20.0&   250\\

B016&           20.0&   165&A136&           20.9&   254\\

C128&           20.7&   165&A049    &       20.6&   259\\

A060&           20.3&   168&C004    &       19.9&   335\\

B164&           20.2&   171&C072    &       20.1&   344\\

C177    &       20.1&   179&A001    &       19.5&   368\\

C015    &       20.3&   186&A282    &       20.6&   385\\
\hline
\end{tabular}
\end{center}

\caption{Variable stars with period estimate. Shortest period objects
  are population II Cepheid-like objects; longest period objects are LPVs.}

\end{table}

\subsection{Period search (time-series analysis)}

Our photometric data, as to time series analysis, are characterised by: - their sampling which is unevenly spaced; - one measurement per night; - yearly season-limited autumn
observations. We are able to detect periods in the
range 60-400 days (and we will see that this range is appropriate for the study of bulge variables which contain an interesting population of long-period objects). 
In this work we do not  extract an extensive sample of
periodic variable stars but rather  identify a subset to characterise our sample.
 We use the Scargle method
\cite*{scargle82} as implemented in MIDAS
for the Time Series Analysis of our data. It is a modification of the
periodogram analysis, equivalent to best-squares fitting of
sine waves to the data and  suited to unevenly spaced data 
\cite{scargle82}. For each variable star, we have fitted  the Scargle
sinusoid on the light curves from the three seasons and made a visual
inspection. A majority of these light curves could correspond to
periodic or semi-periodic objects with periods greater than 100
days. The longest observational season (1995)  spans about
150 days. Periods shorter than 150 days are thus favoured for giving a convincing visual confirmation of the light-curve fit. For periods greater than
250 days, light curves with a sharp feature
which is clearly observed on each year are accepted. For these longest periods, we
checked systematically that the half period did not fit. We present in Fig. 2 to Fig. 10 nine of these
light curves fitted by their Scargle sinusoid (X-axis is in
(J-2449624.5) days, Y-axis in ADU/sec). 

Table 2 gives $R_{max}$ and periods for 54 stars for which the visual fit of the Scargle sine was as good as in the cases of Fig. 2 - Fig. 10. The period estimate is given
with 10 to 20\% accuracy. Since our period determinations are only estimates, we don't give them in our final catalogue.

\subsection {\bf Period Luminosity Diagram}

The  Period Luminosity Diagram is a classical representation for
characterising pulsating variable stars. We thus plot our data
presented in Table 2 in such a diagram in Fig. 11. In order to
characterise the nature of our variable stars we have drawn fiducial
lines for the location of M31 cepheids \cite{freedman90} LMC RV Tauri
stars \cite{alcock98} and Miras \cite{clayton69}, the two last sets of
data being converted to M31 distance. M31 cepheids and LMC RV Tauri
stars have been observed in the $R$ passband and obtaining these lines
was rather straightforward. On the other hand, as for Miras, almost all
recent investigations for P/L studies are made in infrared colours or
are transformed in  bolometric magnitudes. Thus, we decided to use the
$V$ passband absolute magnitude / period diagram of Clayton and Feast
\cite*{clayton69}, which we transformed to $R$ passband at M31
distance. Colour equation from $V$ to $R$ is derived from diagrams from
MACHO data \cite {alves98}. We used also this reference to check the
reliability of the Clayton and Feast's data for long period Miras,
since these authors said there were uncertain.  The astrophysical
aspect of this diagram is discussed in section 5.2.     
 
\section{Variable star general catalogue}

\subsection{The catalogue}

Each variable star in the catalogue is  identified as AGPVm,
m running from 1 to 1579. Then are given: Fn(x,y) (F being the letter of identification of AGAPE field and n the variable star number in this field; x and y being the pixel coordinates in
the reference frame), right ascension and declination for 2000.0, $R_{max}$ 
magnitude at observed maximum as defined in Sec.\ 3.2, and  $L$ the
likelihood
function as defined in section 2.2. Table \ref{table3} presents an extraction of
this catalogue, showing the first and last star for each field and the
12 nova candidates with their numbers added (with respect to 
CDS catalog) in the last column. When the novae were discovered by eye, no $L$ value
is given.

\begin{figure*}

\begin{minipage}[t]{.24\linewidth}
\centering\psfig{figure=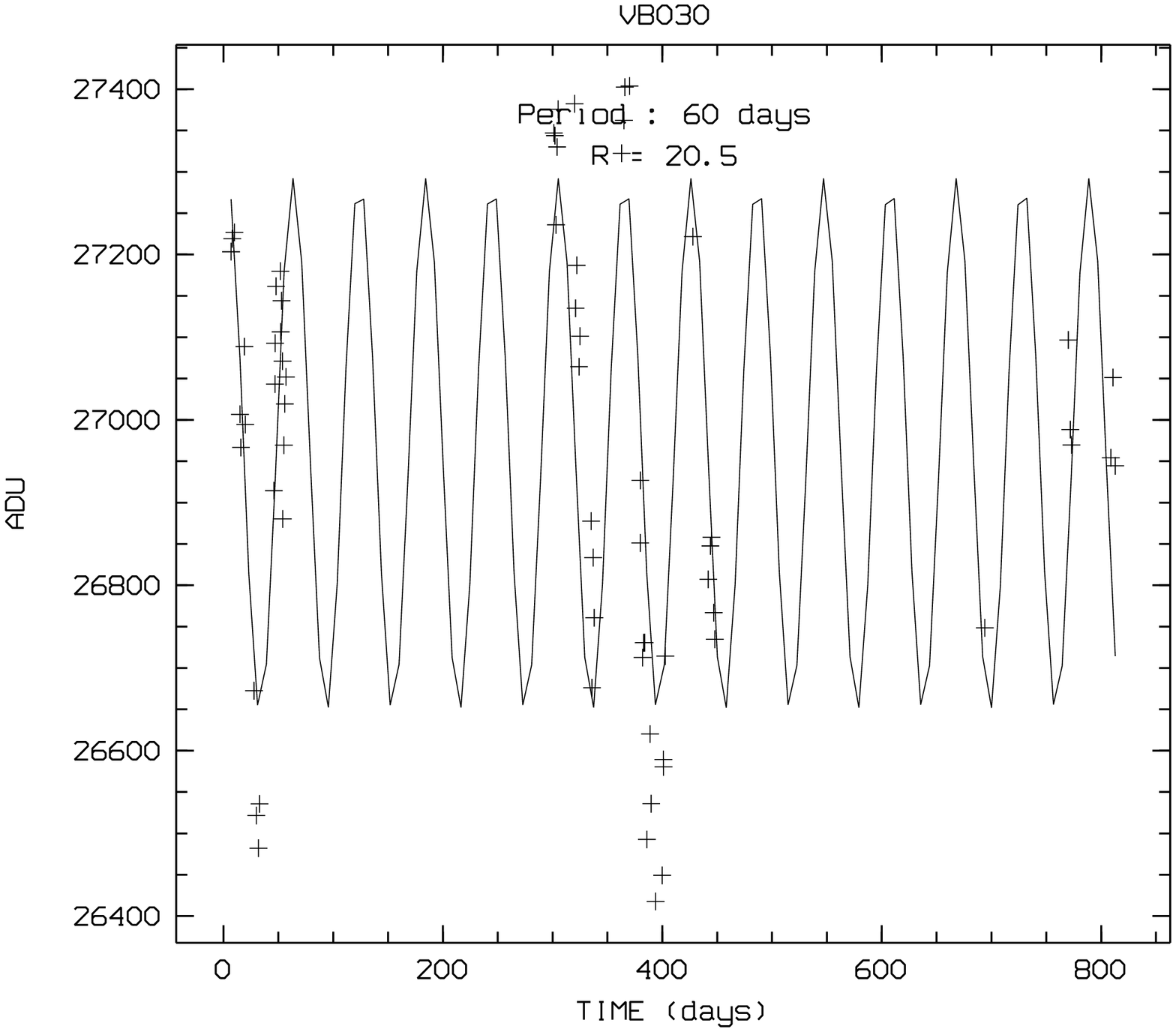,width=\linewidth,angle=-90,clip=}

\caption{Variable star B030: $P$=60\,days ; $R$=20.5}
\label{f01}
\end{minipage}
\hfill
\begin{minipage}[t]{.24\linewidth}
\centering\psfig{figure=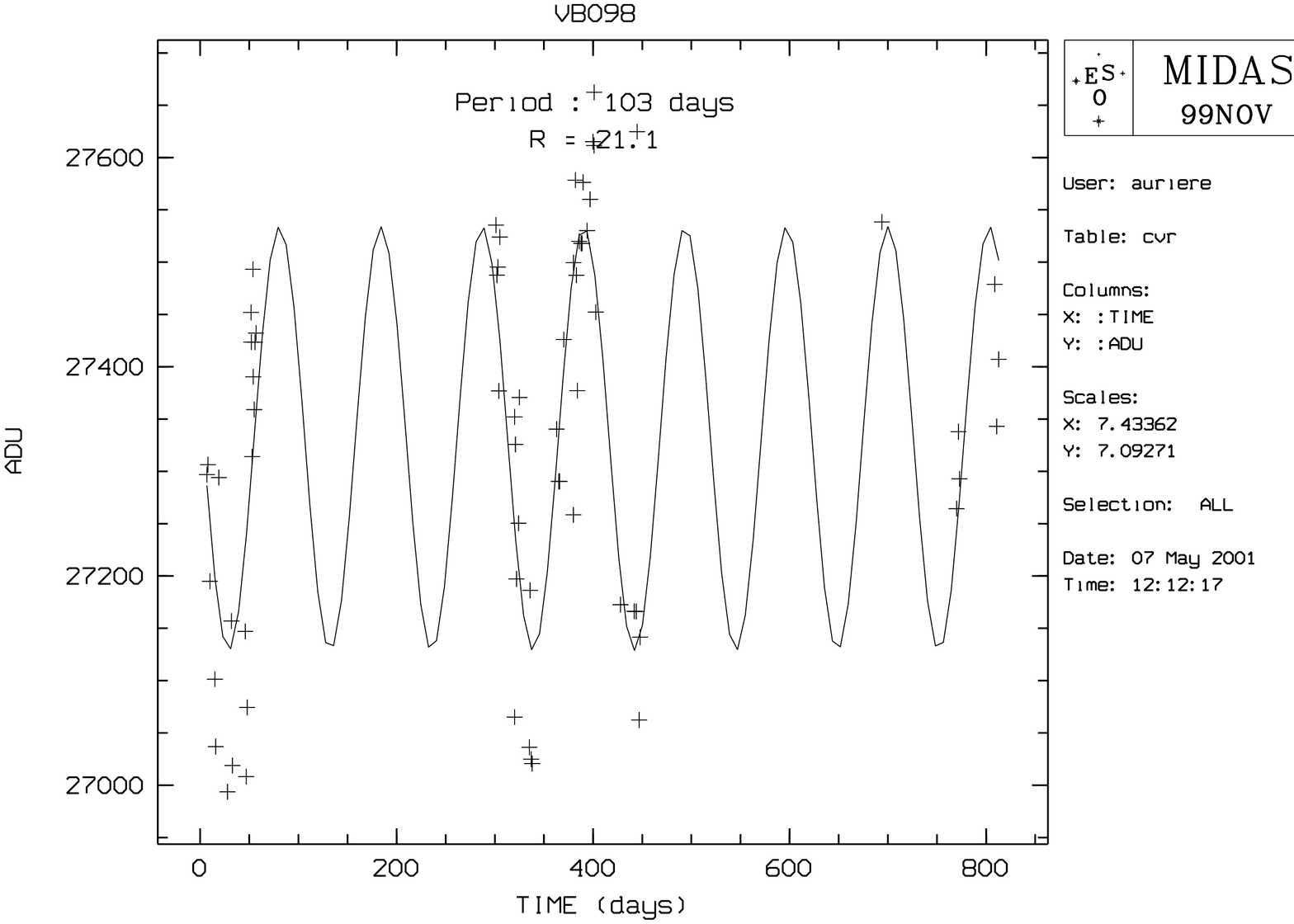,width=\linewidth,angle=-90,clip=}
\caption{Variable star B098: $P$=103\,days ; $R$=21.1}
\label{f02}
\end{minipage}
\hfill
\begin{minipage}[t]{.24\linewidth}
\centering\psfig{figure=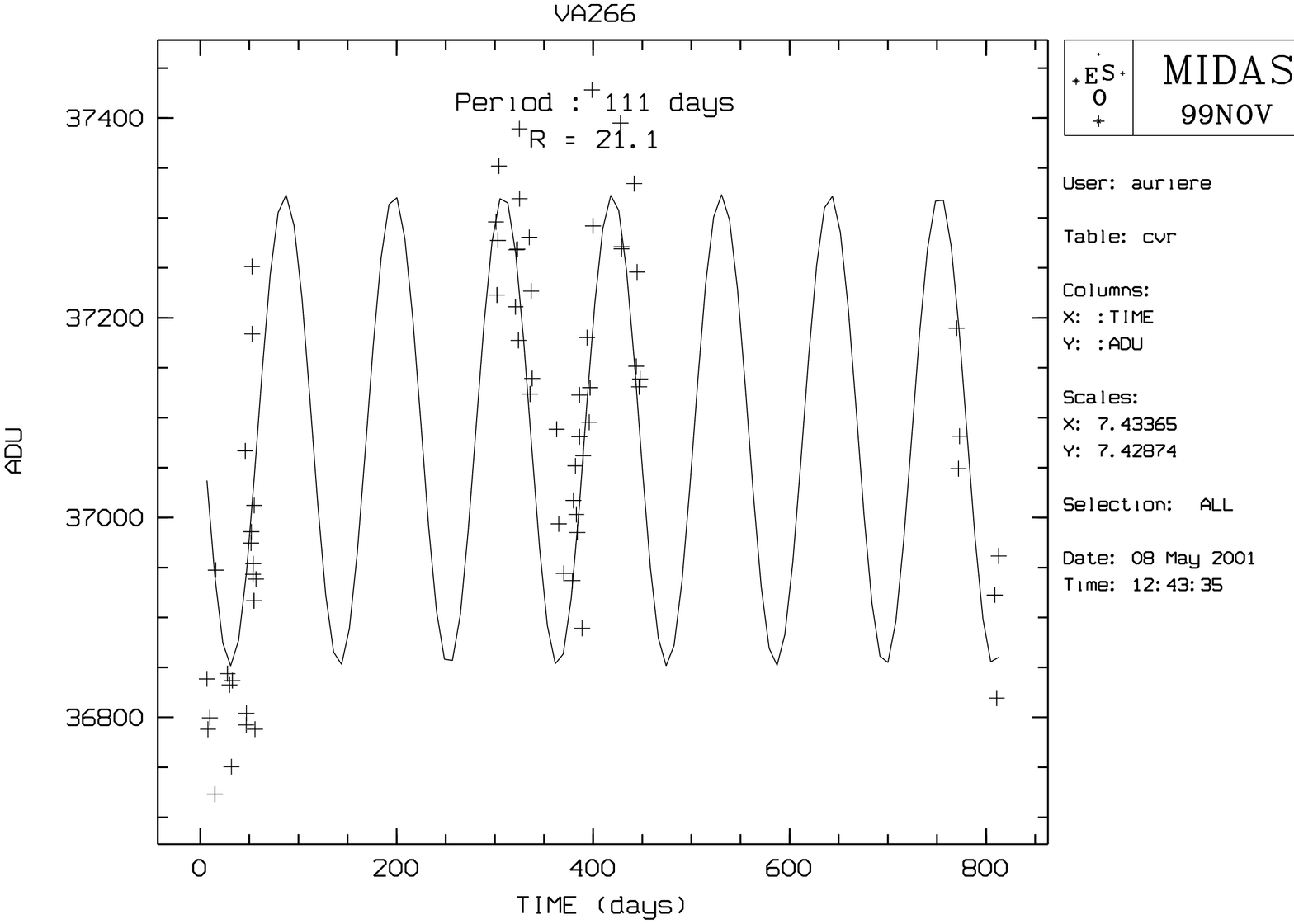,width=\linewidth,angle=-90,clip=}
\caption{Variable star A266: $P$=111\,days ; $R$=21.1}
\label{f03}
\end{minipage}

\begin{minipage}[t]{.24\linewidth}
\centering\psfig{figure=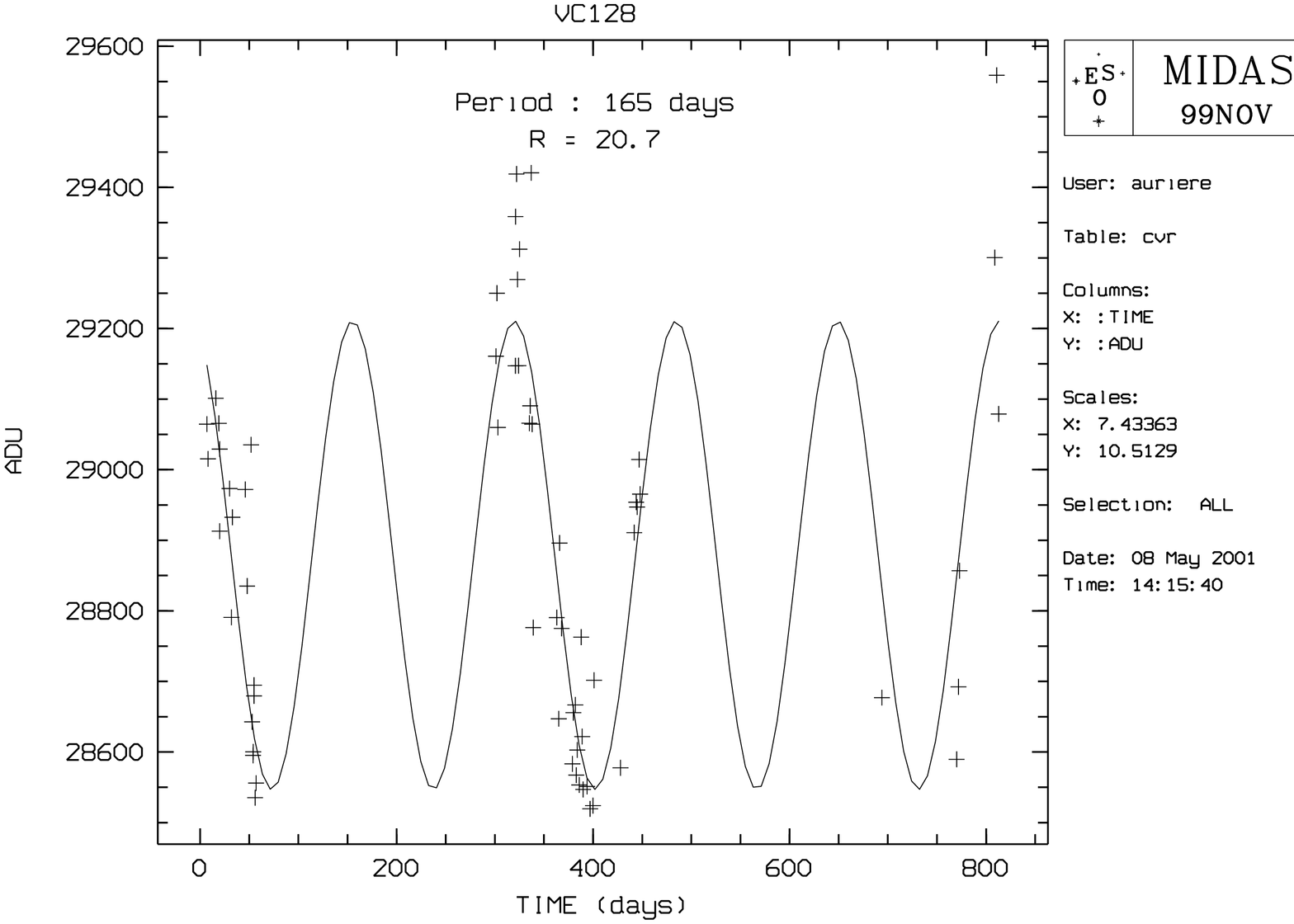,width=\linewidth,angle=-90,clip=}
\caption{Variable star C128: $P$=165\,days ; $R$=20.7}
\label{f04}
\end{minipage}
\hfill
\begin{minipage}[t]{.24\linewidth}
\centering\psfig{figure=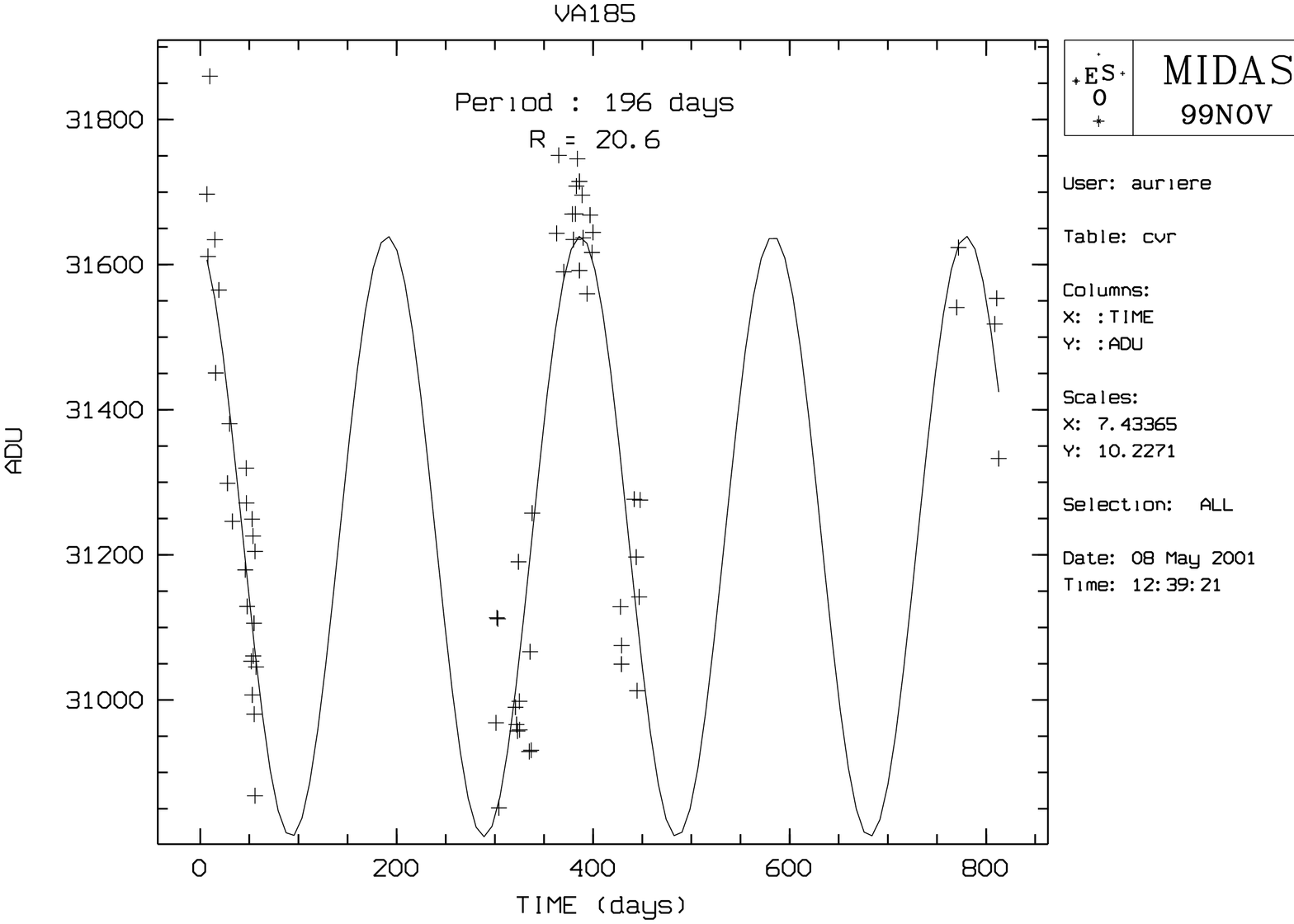,width=\linewidth,angle=-90,clip=}
\caption{Variable star A185: $P$=196\,days ; $R$=20.6}
\label{f05}

\end{minipage}
\hfill
\begin{minipage}[t]{.24\linewidth}
\centering\psfig{figure=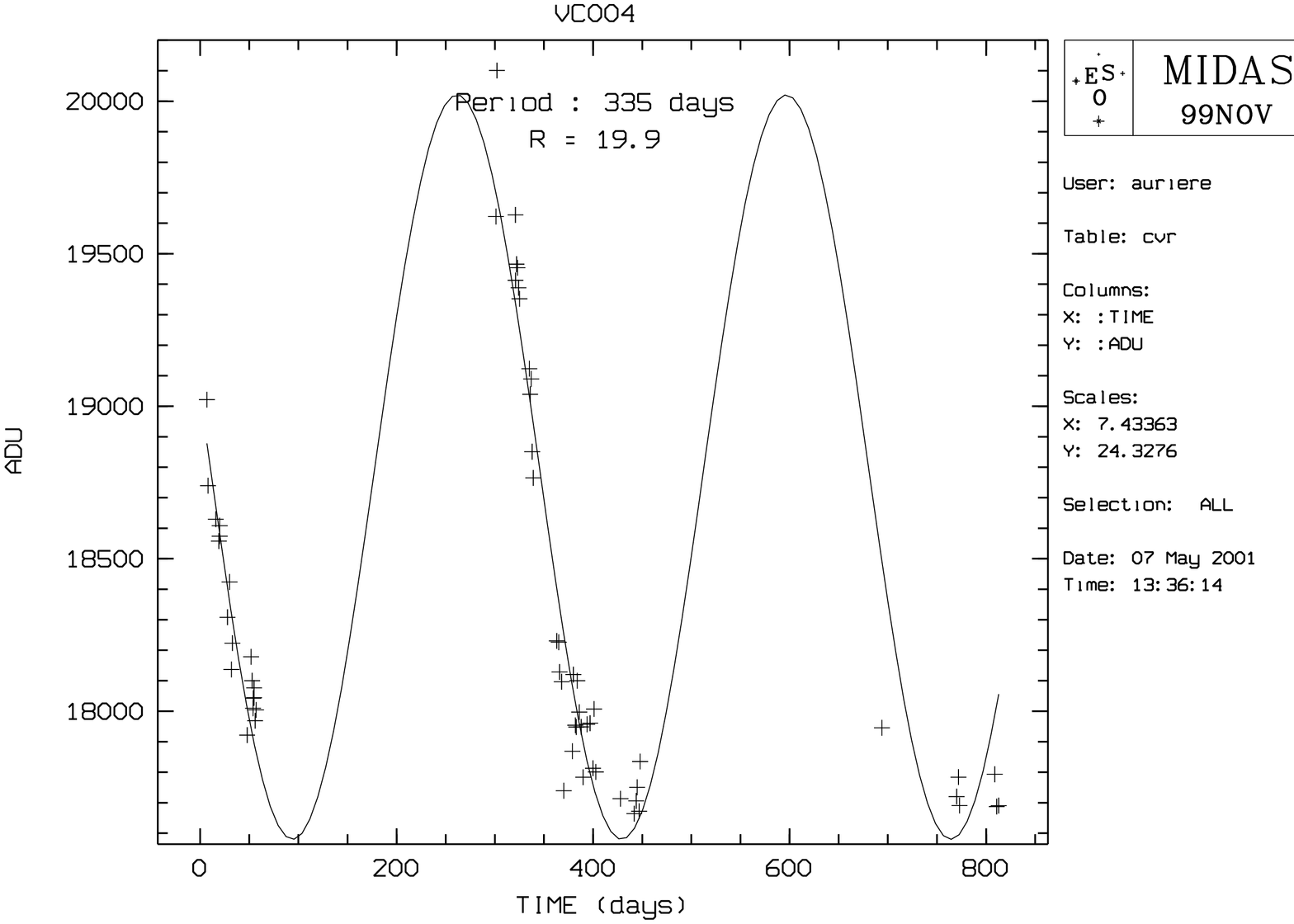,width=\linewidth,angle=-90,clip=}
\caption{Variable star C004: $P$=335\,days ; $R$=19.9}
\label{f07}
\end{minipage}

\begin{minipage}[t]{.24\linewidth}
\centering\psfig{figure=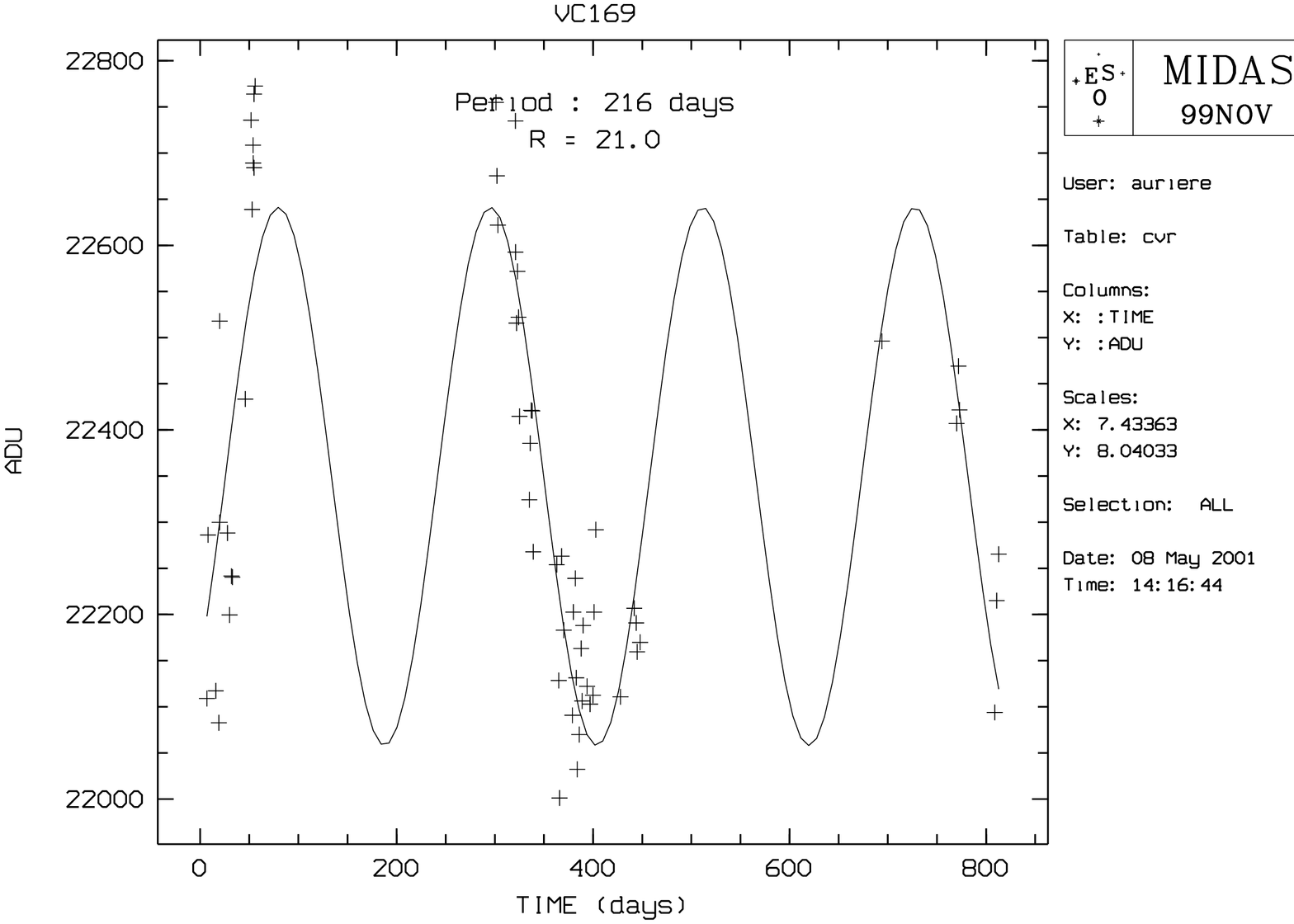,width=\linewidth,angle=-90,clip=}
\caption{Variable star C169: $P$=216\,days ; $R$=21.0}
\label{f06}
\end{minipage}
\hfill
\begin{minipage}[t]{.24\linewidth}
\centering\psfig{figure=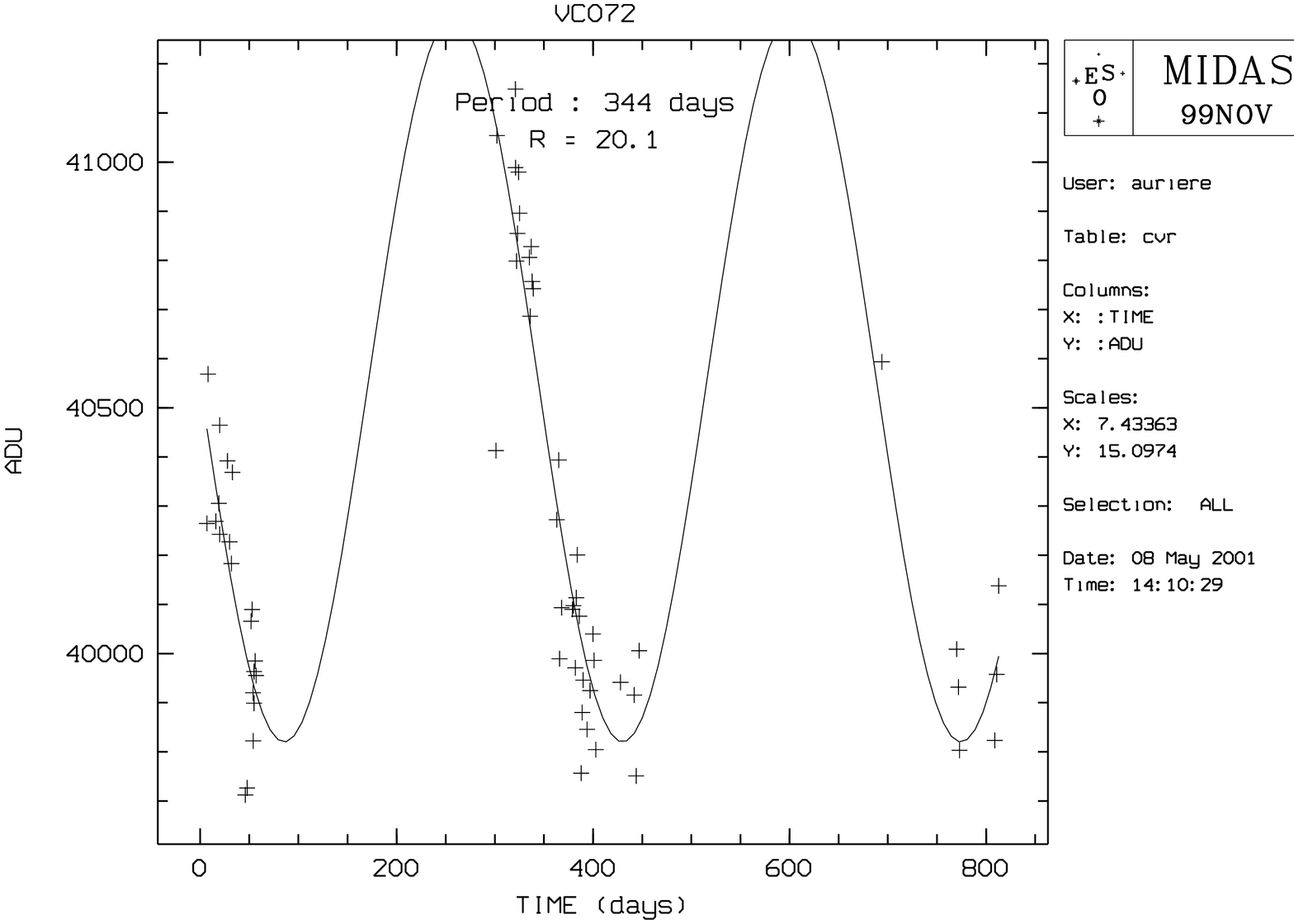,width=\linewidth,angle=-90,clip=}
\caption{Variable star C072: $P$=344\,days ; $R$=20.1}

\label{f08}
\end{minipage}
\hfill
\begin{minipage}[t]{.24\linewidth}
\centering\psfig{figure=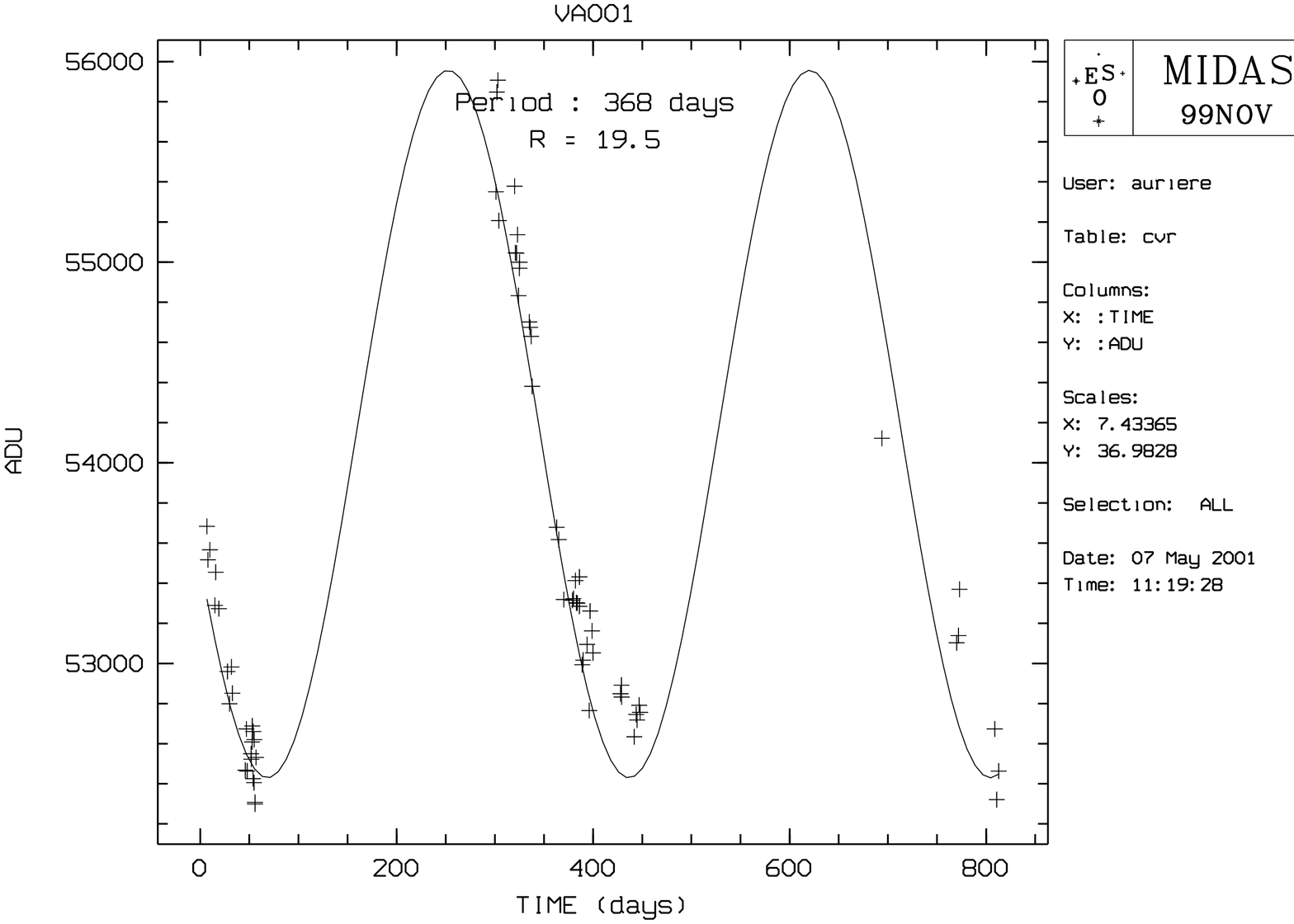,width=\linewidth,angle=-90,clip=}
\caption{Variable star A001: $P$=368\,days ; $R$=19.5}
\label{f09}
\end{minipage}
\hfill

\begin{center}

{\bf Fig.2.-Fig.10.} Variable light-curves fitted by their Scargle
sinusoid. X-axis is in (J-24449624.5) days, Y-axis in ADU/sec. The 3
first objects are of the Cepheid type, the 5 last ones are certainly
LPVs.  
\end{center}
\end{figure*}

\begin{table*}[tb]
\begin{center}
\begin{tabular}{ccccccc}
\hline
\hline
& & & & & & \\[-8pt]
Id  &     Fn$(x,y)$&     $\alpha$(2000.0)& $\delta$(2000.0) &$R_{max}$ & $L$ & \# \\[2pt]
\hline

& & & & & & \\[-8pt]
   AGPV0001 & A001(0697,0392) &  0h42m53.36s &  41d14'26.8"&   19.48 &  940 &\\

\ldots& & &  & & &\\

 AGPV0390 & A397(0834,0352)  & 0h42m50.26s  & 41d14'00.8" &  20.46 &  100 &\\
 AGPV0391 & A398(0365,0380)  & 0h42m58.52s  & 41d15'49.8" &  17.14& &NOV01\\[2pt]

& & & & & & \\[-8pt]
 AGPV0392 & B001(0920,0220) &  0h42m26.41s &  41d16'56.2"  & 16.07& 32000&NOV03\\
AGPV0393 & B002(0754,0333) &  0h42m31.57s  & 41d17'16.0" &  17.24& 20000&NOV02\\
AGPV0394&  B003(0422,0305) &  0h42m36.32s  & 41d18'41.7" &  19.45&   720&\\

\ldots& & & & & &\\

AGPV0581 & B190(0446,0372)  & 0h42m37.40s  & 41d18'23.7"  & 20.94 &  100&\\[2pt]

& & & & & & \\[-8pt]
AGPV0582 & C001(0597,0350) &  0h42m50.65s  & 41d21'20.4" &  19.28&  3000&\\
 AGPV0583 & C002(0784,0384) &  0h42m48.36s  & 41d20'28.5"  & 18.47 & 2800&NOV04\\

\ldots && &  & & &\\
 AGPV0842 & C272(0771,0504) &  0h42m51.18s &  41d20'09.8" &  21.11  & 100&\\[2pt]

 & & & & & & \\[-8pt]
 AGPV0843 & D001(0910,0609) &  0h43m 8.41s &  41d16'39.9" &  16.80& 30000&NOV05\\
 AGPV0844 & D002(0927,0567)  & 0h43m 7.22s &  41d16'43.4" &  19.39&  1800&\\

\ldots & & & & & &\\
AGPV1185 & D348(0585,0580)  & 0h43m13.04s  & 41d18'04.9"  & 21.43 &  100&\\[2pt]
 & & & & & &\\[-8pt]

AGPV1186 & E001(0721,0300) &  0h42m17.24s  & 41d13'35.4" &  17.65 & 4900&NOV08\\
 AGPV1187 & E002(0459,0646) &  0h42m29.00s  & 41d13'36.1" &  18.22& &NOV06\\
 AGPV1188&  E003(0317,0264) &  0h42m23.07s  & 41d15'20.6" &  20.63  & 490&\\
\ldots & & & & & &\\
AGPV1438 & E259(0337,0392) &  0h42m25.51s  & 41d14'52.3" &  21.00  & 100&\\
AGPV1439 & E260(0724,0490) &  0h42m21.30s  & 41d20'28.7" &  18.41& & NOV07\\[2pt]

& & & & & & \\[-8pt]
AGPV1440 & F001(0905,0390) &  0h42m33.27s  & 41d09'59.6" &  20.35 &  520&\\

\ldots & & & & & &\\
 AGPV1575 & F138(0425,0400) &  0h42m41.22s  & 41d11'55.2" &  20.89  & 100&\\
 AGPV1576 & F139(0346,0381) &  Oh42m42.08s &  41d12'18.0" &  17.56&&NOV09\\
 AGPV1577 & F140(0430,0650) &  0h42m46.60s &  41d11' 8.3" &  17.82&&NOV10\\[2pt]

& & & & & & \\[-8pt]
AGPV1578 & Z001(0528,0266) &  0h42m43.75s &  41d16'52.0"  & 16.35&&NOV11\\
AGPV1579 & Z002(0658,0411)  & 0h42m44.81s  & 41d15'53.6" &  14.98&&NOV12\\[2pt]
\hline

\end{tabular}

\caption{Extraction of the catalogue of the AGAPE variable stars 
showing the first and last star for each field and the 12 novae
candidates (more information are given in the text). $L$ is the
likelihood function the nova number added (with respect to CDS 
catalog) in the last column; more information on the columns is given in
sect. 5.1 \label{table3}}
\end{center}
\end{table*}

\subsection{Correlation with other catalogues}

We have then correlated our catalogue with the General Catalogue of
Variable Stars (GCVS, Durlevich et al. \cite*{durlevich96}) as available at CDS, with Chandra sources \cite{kaaret02,kong02,distefano03}, as well as 
2MASS data \cite{skrutskie97}. As to the GCVS, about 200 objects fall in our field, all being
classified as novae but one cepheid. 7 objects of GCVS (all novae) fall at less than 2" from our variable stars. Among these associations, only one concerns one of our nova candidates and we consider that the other ones are chance associations. Nova 27 from GCVS is at 1.9" from NOV12 (AGPV1579, Z002, sect. 5.3). Nova 27 was observed in 1985 \cite{ciardullo87} and could then be a recurrent nova. 

M31 V0934 of GCVS, which is V38 of  \cite{hubble29}, is a type I cepheid (period 17.764 days). It falls in our field C but has no counterpart in our catalogue. However, we find on our data one variable object located 2.5" from V0934 position with $R$ about 20 magnitude at maximum. M31 V0934 may have been discarded in our microlensing survey as being one resolved star on our reference frame (sect. 2.2).

Chandra made several observations of the bulge of M31. Tens of X-ray
sources were detected with location accuracy better than 1" (Kaaret
\cite*{kaaret02}, Kong et al. \cite*{kong02}). These X-ray
observations are the only ones which deserve correlation with our
catalogue since they have small error boxes (1" accuracy versus
several several " for ROSAT HRI) necessary to avoid too many chance
identifications. Among the 4 X-ray sources which fall nearer than 2"
from our variable objects, 2 deserve peculiar attention. CXOM31
J004242.1+411218  \cite{kaaret02} is located 0.7" from NOV09
(AGPV1607, F139). This identification is discussed in sect. 5.3. 

CXOM31 J004247.8+411550. is located 0.6" from AGPV0288, in the very central bulge. Kong et al. \cite*{kong02} observed it again as r1-25. Di Stefano et al \cite*{distefano03} include 
this source in their list of super soft X-ray sources candidates. Di stefano et al. \cite*{distefano03} could not find any optical counterpart brighter than magnitude 21 on HST archives, in the $R$ wavelenth range. AGPV0288 peaks at $R=19.7$ in AGAPE observations, and is a small amplitude irregular object. r1-25 is fitted with a blackbody model kT=122\,eV, rather hard for a supersoft source. Clearly, this association will deserve a deeper study.

About 50 objects from 2MASS catalogue fall in our field. None  are associated with AGAPE variable stars within 2" radius, but some are associated with brighter objects (galactic stars or globular clusters). This is
in accordance with the fact that the rather short period LPVs of our
catalog are expected to be fainter in K that the limiting
magnitude of  the 2MASS catalog (about K=16), when scaling LMC
results \cite{hughes90,lebzelter02}.

\subsection{Spatial distribution of variable stars and the extinction
  in M31 bulge} 

The spatial distribution of the variable stars is of importance for
two reasons : variable stars are often considered as a possible
reference background  from which it will be possible to disentangle a
microlensing event population which will present a near/far M31
asymmetry \cite{crottsm31,kerins01};  part of the variables we have observed
really belong to the bulge and we can thus contribute to the disk/halo
distribution debate of objects in M31. An important factor in the
star distribution in M31 comes from extinction \cite{hatano97}.

In the AGAPE survey, the six investigated fields are distributed apart
from the major axis of M31, A D F being on the ``far side'' and B C E on
the ``near side''. We consider that the E and F fields have too sparse
observations to give significant data (cf. Table 1) for this
problem. At the near side, which corresponds to the ``bottom'' of \cite{hatano97}, the bulge is in part behind the (dusty) disk. In our
study, the most significant anomaly in the variable star distribution is
due to B field. It contains half the number of variable stars in field
A (190 with respect to 370; almost same ratio for the corresponding
numbers of stars per brightness unit). The B field contains extinction
zones such as the dust complex D395A  \cite{melchior00b} . However, using
the extinction map of Melchior et al. \cite*{melchior00b}, we did not find a
correlation between the variable stars and extinction
distributions. This could come because extinction can be  very patchy
in galaxies \cite{thoraval99}. Local large extinction has been pointed
out in M31 in the case of globular clusters \cite{barmby02}. 

Thus, because of extinction and perhaps threshold effects, the variable
star distribution towards the bulge of M31, might be not symmetric with
respect to the major axis of the galaxy.

\section{Identified variables}

\subsection{Introduction}

Our observational data enable us to characterise  variable stars with
their absolute magnitudes, colours and periods. Observational
conditions of the AGAPE experiment bring some biases and thus some limitations
 on the
identification of the variable stars which were discussed in previous sect.s.  Our robust variability criteria lead to almost all measured $R_{max}$
of variable stars being between 20 and 21 magnitudes. Inspection of tables of data published by the DIRECT collaboration
shows that the bulk of their detected variable stars is
in the same magnitude range as ours. 

Our search  detects M31 variable stars brighter than 
$R=21$ or $M_R<3.3$. 
Colour magnitude diagrams such as that of Kukarkin \cite*{kukarkin75}
show that this limit enables the study of only the brightest part of the
variable star population including long period
red stars and Cepheids, and almost
all the kinds of stars observed by Baade et al. \cite*{baade63,baade65}. On the other hand, RR Lyrae
stars, which are difficult to observe from the ground in M31 because of their
faintness
\cite{pritchet87}, are clearly out of reach. 

We are able to study the population II variable
stars which are tracers of the bulge in a region where by projection
bulge and disk populations are not easy to disentangle. Few population
II objects like RV Tauri stars have been studied, and  Miras have not
been unambiguously detected in M31. Up to now variable surveys were
done more in the disk population than in the bulge. Studies have shown
that when approaching the center of M31 a number of new star classes appeared. This is
first the case for bright red stars, some of them being possibly very
long period Miras \cite{rich91}. Hodge \cite*{hodge92} remarks that
from the work of Swope and Baade \cite*{baade63,baade65} the number of  
irregular variables
increases when approaching the bulge. Also, the same author shows that
the ratio of type I Cepheids to type II Cepheids has a maximum in
medium parts of the disk. Thus one expects to find very few type I
Cepheids in the bulge direction.
Apart from novae which are outstanding bright objects  in our sample,
other objects can be characterised by the CMD of the population
resolved in $B$. Figure 1 shows that this population is peaked on
$R$ around $20.4 \pm  0.5$ mag. ($M_R = -4 \pm 0.5$) and $B-R$ around
$1.7 \pm 0.5$ mag. The photometric properties of the variable stars
when compared to the HR diagram of Kukarkin \cite*{kukarkin75}, are compatible
with faint RV Tauri variables or bright semi-regular ones. The bulk of
the measured periods is in the range 100-250 days and thus they are
compatible with this hypothesis. It is also remarkable that the bulk
of the CMD of Fig. 1 is very similar (taking into account the distance
modulii) to that obtained by Melchior et al. \cite*{melchior00} for
variables detected using the pixel method in the LMC, which are
interpreted as ''Long Timescale and Long Period Variables''. 

In the two  following sections, we describe work
on variable stars for which we could identify the type, namely rather long period pulsating objects and novae.

\begin{figure}[htb!]
\begin{center}
\psfig{file=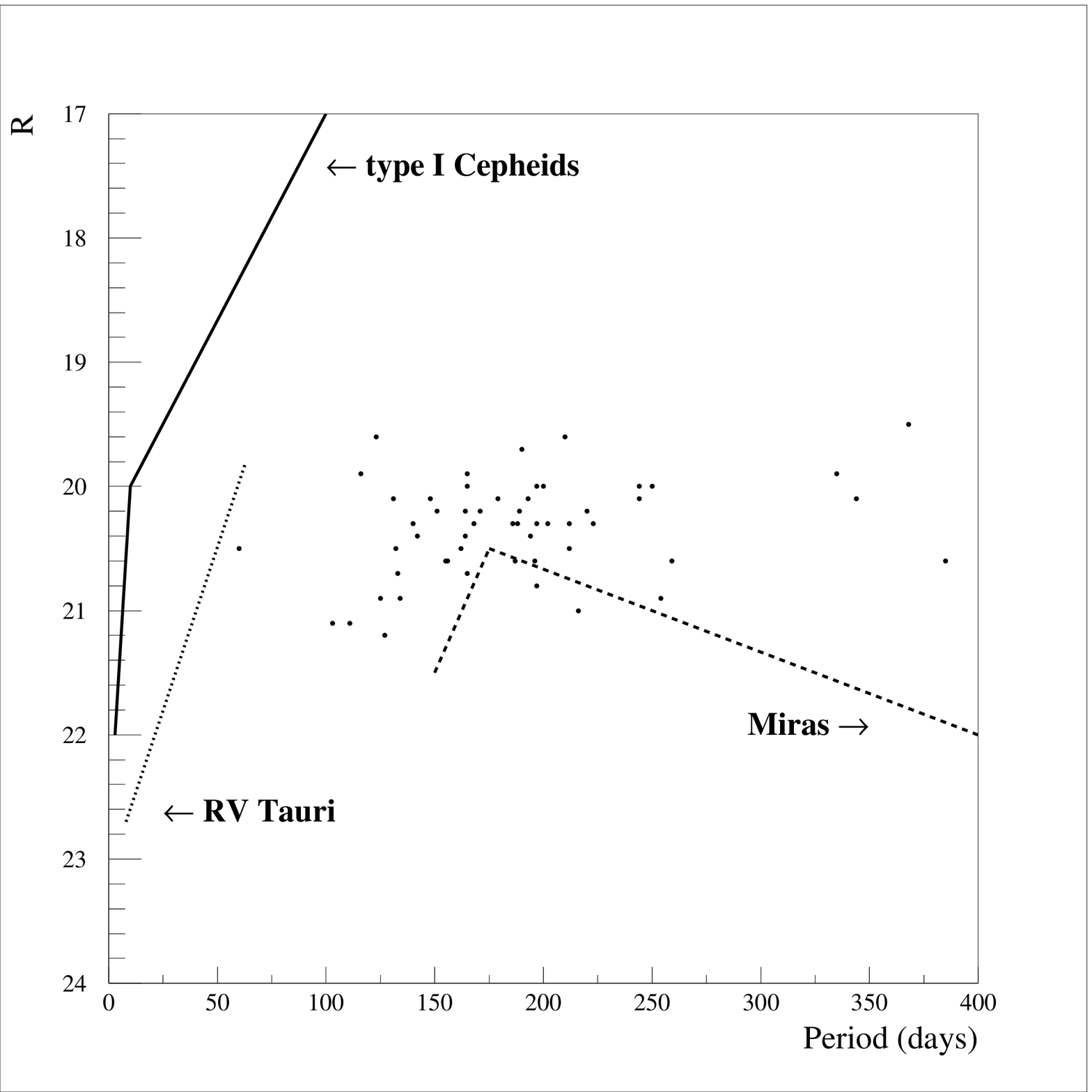,,width=.4\textwidth}
\caption{Period Luminosity Diagram: $R$ magnitude is plotted against the
  period for the stars selected in Table2. Locations for Cepheids, RV
  Tauri stars and Miras are shown. Details are given in the text.}
\label{fig11}
\end{center}
\end{figure}

\subsection{Periodic pulsating stars}

Bright pulsating stars are generally divided into Cepheid-like and
long period red 
variables (called hereafter LPVs). In the part of the colour magnitude diagram
we investigate here, as well as in the 100-200 days period interval,
these two kinds of variable stars overlap. However, since LPVs
have generally periods greater than 100 days, we can expect the
shortest periods variable stars in our catalogue  to be Cepheid-like
(example Fig. 2). Actually, some  of the light curves are
reminiscent of the asymmetric shape observed for cepheids
\cite{stetson96}. On the other hand, stars with periods greater than 200 days are presumably
LPVs (examples Fig. 8, 9, 10).

Early work on M31 Cepheids was done by Baade and Swope
\cite*{baade63,baade65} with the Palomar telescope to calibrate the M31
distance. Modern CCD studies were made  \cite{freedman90,magnier97,mochejska00} and the period luminosity for M31 population I Cepheids
is by now well established for several colours including $R$. 
Figure 11 shows that our shorter period variable stars are at least 2
  magnitudes fainter than M31 population I cepheids
  \cite*{freedman90}. They thus clearly belong 
to the population II type \cite{wallerstein84}. Actually, the shortest period star of Table 2 lies very near the location of LMC RV Tauri stars, for which, unfortunatelly the long period range is not known \cite{alcock98}. The light-curve of one star of this type in M31 has been recently presented (Riffeser et al, 2001) ; however, our data sampling is not good enough to uncover the characteristic double-wave shape of RV Tauri stars in our light curves.

LPV (Period$ >$ 100 days) are classically divided into
Miras and semi-regular variables \cite{habbing96}; the Miras
are classically those with amplitude larger than 2.5 mag. Microlensing surveys have brought an impressive
number of new variables, infrared  data (both near-infrared from the
ground and far-infrared from space) Hipparcos and HST observations
have completely renewed our knowledge on pulsating stars and the two
classes are now considered  as overlapping \cite{bedding98,lebzelter02}. Optical light curves and optical
colours of  LPVs have instigated new studies
\cite{delaverny97,mennessier97}. Figure 11 shows that our stars are
  generally brighter than what is expected for a Mira \cite {clayton69}.
Since semi-regular variables are rather brighter than Miras, they should be the more
frequent in our catalog and Table 2. Yearly windowing has certainly
biased our selection towards near-one-year periods. We checked that
half the derived periods did not give plausible fits. C004 (Fig. 8), C072
(Fig. 9) and A001 (Fig. 10) are particularly convincing. A001 presents a 3
magnitude amplitude light curve and is very likely to be a
Mira. However, these objects are more than one magnitude brighter than
what is  expected for galactic or LMC Miras with the same period. We cannot exclude neither peculiar LPVs nor that our phase undersampling leaded us to a false (too long) period determination.

\begin{table}[hbt]
\begin{center}
\begin{tabular}{ccccc}
\hline
\hline
 \# &  Id& $R_\mathrm{max}$ &  Date &  year \\

\hline
 NOV01 & AGPV0391 & 17.14 & 006.88 & 1995\\

 NOV02  &   AGPV0393&17.24  &399.83  & 1995\\ 

 NOV03  &      AGPV0392&  16.07 &303.10& 1995\\

 NOV04  &      AGPV0583&  18.47 &771.89& 1996\\

 NOV05  &     AGPV0843& 16.80&335.14& 1995\\

 NOV06 &       AGPV1187&  18.22 &301.12& 1995\\

 NOV07  &     AGPV1439& 18.41&006.96&  1994\\

 NOV08 &      AGPV1186& 17.65&443.93 &  1995\\

 NOV09 &        AGPV1576& 17.56&006.97& 1994\\

 NOV10  &       AGPV1577&17.82 &321.10& 1995\\

 NOV11  &       AGPV1578&  16.35&320.94& 1995\\

 NOV12  &    AGPV1579& 14.98 &444.82& 1995\\
\hline
\end{tabular}
\caption{\label{tab4}Characteristics of the 12 novae candidates named as NOVn (first column) where n goes
from 01 to 12.}
\end{center}

\end{table}

\subsection{Novae}

M31 novae have been systematically searched for more than 70 years 
\cite{hubble29,arp56,rosino89,ciardullo87,shafter01}. 
The relation between their magnitude and rate of
decline is now established \cite{capaccioli89} and
can be compared to those of other galaxies \cite{dellavalle95}. 
In addition to being interesting in their own right, novae are of cosmological
interest in the context of Type Ia supernova precursors and the nature of
supersoft sources \cite{kahabka97}.

Twelve nova candidates were detected by  AGAPE. These objects are so
bright at maximum that some were found visually and we can
expect detection of all novae which occurred in our field during AGAPE
observation periods and for which 3 points of the light curve were measured. 

\begin{figure*}[p]
\psfig{file=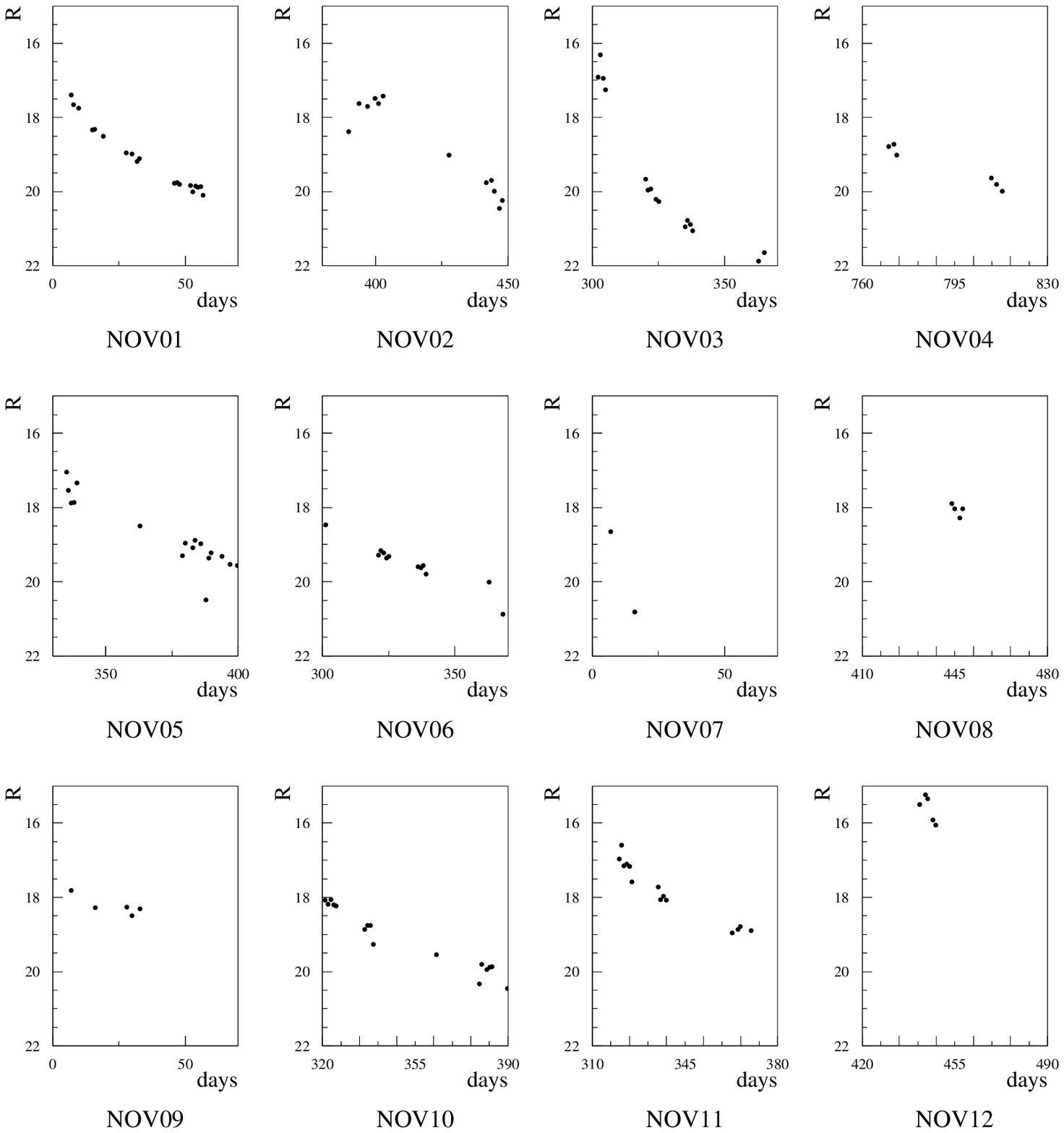}

\caption{Light curves of the 12 novae candidates detected in the Pic-du Midi AGAPE survey}
\label{fig12novae}
\end{figure*}

Table~\ref{tab4} gives the data for our 12 nova candidates named as NOVn (first column) where n goes
from 01 to 12. Column 2 gives the catalogue identification, so that coordinates can be found in Table 3 (see also section 5.1). The next three columns give $R_{max}$, the date of the observed maximum brightness given in (J-2449624.5) days, and the year of occurrence. Figure 12 shows the light curves for these twelve nova candidates. Only six
of the novae were observed at maximum. NOV02, 04, 06, 08, 11
were observed by Shafter and Irby \cite*{shafter01}. NOV08 and NOV09
present very small brightness decreases and are peaked around  $R=18$.
They could correspond to the wing part of bright nova light-curves, as
NOV06 could be as well. Actually AGAPE observed NOV08 during 1995's
season and Shafter and Irby  \cite*{shafter01} in 1996, when it was the
brightest object of their survey. NOV08 could thus be a recurrent
nova. Also one nova appears to be observed at 1.9" from NOV12 by Ciardullo et al. \cite*{ciardullo87} in 1985 (their nova 27, see sect. 5.1). NOV12 could thus be a recurrent nova. 

Now, the Chandra source CXOM31 J004242.1+411218. \cite{kaaret02} is
located 0.7" from NOV09 (see sect. 5.1). NOV09 was observed in 1994's
AGAPE runs and Chandra observations were performed in 2001
october. Orio et al. \cite*{orio01} describe which can be expected as
X-ray emission from classical and recurrent novae. Most classical and
recurrent novae emit hard X-rays in the first months after the
outburst and during 2-10 years, with peak X-ray luminosity of a few
times $10^{33}$ erg/s. On the other hand, the supersoft X-ray phase of
novae which is also relatively short lived (up to 10 years) is
observed only for up to 20\% of novae according to 108 objects
analysed contained in the ROSAT archive. This phase corresponds to
very soft spectra (equivalent temperatures in the range 15-80 eV) and
high luminosity ($10^{36}$ - $10^{38}$ erg/s bolometric luminosity)
\cite{kahabka97}. CXOM31 J004242.1+411218. has a luminosity of 
$0.4\;10^{36}$ erg/s and thus cannot be associated with the hard emission of
an M31 nova. On the other hand, scaling an estimate presented by
Kaaret \cite*{kaaret02}, a blackbody spectrum with a temperature of 80
eV would imply an unabsorbed luminosity of $1.2\;10^{36}$ erg/s in the
0.1-10 keV band for HRC count rate of this source. Lower temperatures
would require higher unabsorbed luminosities to produce the same
observed count rate. The luminosity of CXOM31 J004242.1+411218 is thus
consistent with that of a supersoft source. Unfortunately it is a
rather faint source and no spectral information is available. No other
observations with Chandra \cite{kong02} nor XMM Newton \cite
{distefano03} report this source. It may be because it is near the
limiting sensitivity of these observatories, but supersoft sources are
also known to be transient sources \cite {trudolyubov01}. Up to now
only 2 X-ray sources were suggested has being possible supersoft ones
linked with M31 nova outbursts (Nedialkov et al. \cite*{nedialkov02},
Kaaret \cite*{kaaret02},Di Stefano et al.\cite* {distefano03}. In
sect. 5.1 we discuss another possible identification of a supersoft
source with an AGAPE catalogue variable object. 

An important question concerning novae is about their
location; both their radial distribution and population to which they belong
is still in debate. First Ciardullo
\cite*{ciardullo87} showed that there is not a central hole in their
distribution as could be derived from \cite{arp56} observations. The debate is to know whether novae belong to the bulge or to the disk.
Ciardullo et al. \cite*{ciardullo87} concluded that the
novae were a bulge population and suggested that the binary central
excess in M31 (novae, X-ray sources) could come from ejection from
globular clusters. Auri\`ere et al. \cite*{auriere92} argued that actually all X-ray sources and novae in the
bulge of M31 were sufficiently near globular clusters that they
could have been ejected from. Shafter and Irby \cite*{shafter01} also support a
main bulge origin though they cannot exclude a bulge fraction as low
as 50\%. On the other hand, Hatano et al. \cite*{hatano97} propose that the M31 novae
belong mainly to the disk population. They made a study of the
positions 
of 191 novae
observed at Mount Wilson and Asiago for which $B$ photometry at
maximum was available, and used a model of extinction in order to interpret asymmetry along the minor axis.
Studies of other systems show
that young populations may be more efficient than old populations to
produce novae. Now, it is classically considered that
disk dominated galaxies produce more novae, which are brighter and
with faster decline, than bulge dominated galaxies which produce
fainter and slower ones \cite{dellavalle94,dellavalle98}. 
This scenario has population-synthesis code support \cite{yungelson97}.  
Our sample of 12 novae candidates does not show any asymetric distribution along the minor axis of M31. However, it is to small to bring a significant contribution to the debate and it will be useful when joined to those from other M31 surveys.

\section{Conclusions}

We have presented the first catalogue of variable stars located in (or
at least towards) the bulge of M31. The 1579 objects include 12 nova
candidates, and mainly semi-periodic or periodic objects. Their
magnitudes, colours and long period of variation suggest that our
sample is dominated by a mixture of Long Period Variables (LPV) 
and type 2 Cepheid like variables (RV
Tauri). The AGAPE data thus can
address the question of the M31 bulge variable 
star population and particularly the nature of its bright component as well as its
spatial distribution. 

It is worth mentioning that both 
novae and the pulsating periodic variable stars are of cosmological
interest as luminosity candles and distance calibrators. Also, the nova population is related to supersoft X-ray sources which may be progenitors of Type I supernovae. As to this aspect, we propose both the detection of two recurrent novae and the possible optical identification of two supersoft X-ray sources candidates observed with Chandra. This work
will be complemented and extended using data from the new large field
survey POINT-AGAPE performed with the Isaac Newton Telescope \cite{an04}.  

{\bf Acknowledgements:}

This research has made use of databases operated by CDS, Strasbourg, France.

\bibliography{agapevar}

\end{document}